# Light Amplification by Active Meta-mirrors

Ignas Lukosiunas[1], and Kestutis Staliunas[1,2,3,*]

[1]Vilnius University, Faculty of Physics, Laser Research Center, Sauletekio Ave. 10, Vilnius, Lithuania
[2]ICREA, Passeig Lluís Companys 23, 08010, Barcelona, Spain
[3]UPC, Dep. de Fisica, Rambla Sant Nebridi 22, 08222, Terrassa (Barcelona) Spain

*Corresponding author: kestutis.staliunas@icrea.cat

**Abstract:** We consider a scheme of thin films, deposited on periodically modulated amplifying materials. We show that the reflection from such meta-interface can undergo substantial amplification, due to Fano resonances in the thin films. The amplification strongly increases when the Fano waveguiding modes approach the edge of the continuum and degenerate into the leaky surface modes. The study is based on simplified analytical models, as well as on Rigorous Coupled Wave Analysis.

**Introduction:** Wave dynamics in the systems of periodically modulated thin films is the subject of increasing interest [1-10]. The coupling between the incoming/transmitted/reflected modes (the Fabry-Perot modes of the film-resonator), and the waveguiding modes of the thin film occurs due to the periodic modulation of the thin film surface(s), see Fig.1. As detailed below in this article, the reflection vanishes, and reaches its maximum at two, close one to other frequencies, thus building a characteristic asymmetric shape of the Fano resonances [11-13]. Different physical ideas have already been realized with such thin films: spatial filtering [14-17], polarizers [18-19], and unidirectional light trapping in non-Hermitically modulated thin films [20]. Recently the study has been extended to the resonances when the waveguiding mode degenerates to the leaky surface modes, where a sharpening of the resonances has been predicted [21].

The studies are performed mainly by numerical calculations, applying Finite Difference Time Domain (FDTD), and Rigorous Coupled Wave Analysis (RCWA). All previous studies concern passive systems. These include dielectric structures, but also sometimes metallic structures, where plasmonic resonances can take place as well [20-21]. Here we explore another idea with the thin films, never addressed before: thin films on the amplifying substrates. The natural expectation is that possibly the radiation, reflected from such a thin film, can experience strong amplification. The idea is easy to understand intuitively: the waveguiding modes are not completely confined in the waveguiding (passive) structure, but leak to both sides – the air (superstrate) from one side, and the substrate, from the other side. The substrate as an amplifying media provides some gain to such waveguide modes, thus enhancing both transmitted and reflected waves. The leakage of the waveguiding mode into the substrate here plays an important role: the more the mode leaks into the amplifying substrate – the more gain it could gather. The extreme case occurs at the edge of the continuum when the waveguiding mode leaks deep into the substrate, i,e. degenerates into the leaky surface mode. The amplification then is expected to immensely increase.

The technical concept of the guided mode resonant amplifiers is shown in Fig.1.

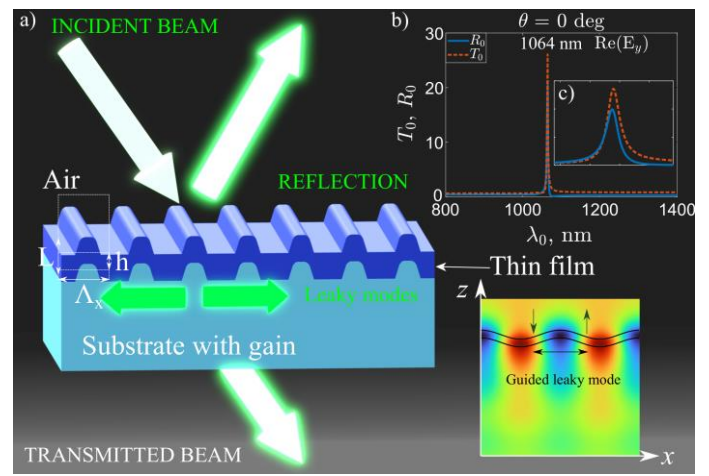

***Fig.1.*** *The concept of an active meta-mirror: a) the incident beam couples into the waveguiding mode of the thin film. Back-coupling results in substantial increase of the reflected ant transmitted radiation. b). Reflection/transmission dependence on the wavelength by crossing the resonance. c) the field distribution in the thin film.*

The article analyzes the effect. The analysis is basically performed by the Rigorous Coupled Wave Analysis (RCWA) numerical model.

**Analytical approach**: The radiation incident at a near-to-normal direction to the film, excites planar modes of the film, with the refraction index larger than that of the environment. Such planar modes can be bounded [17], but also leaky [22]. In order to couple these planar with incident wave, the surfaces of the film are periodically structured on a wavelength scale along its surface, as shown in Fig.1.a. If the planar structure supports the waveguiding modes with the propagation wavenumber $k_m$ for a given frequency $\omega_0 = ck_0$ (here the wavenumber $k_0 = 2\pi/\lambda$), and if the period of the film modulation is $d_x$ (the reciprocal modulation wavenumber is $q_x = 2\pi/d_x$), then the resonant coupling to the right/left propagating modes occurs for the incidence angles $\theta$ such that $k_0 \sin(\theta) \pm q_x = \pm k_m$.

The propagation wavenumbers of the planar modes $k_m$, do not have explicit analytic expressions in general, however in the limit of strongly confining planar waveguide, corresponding to tight binding approximation, the mode propagation numbers can be estimated as $k_m = m\sqrt{(nk_0/m)^2 - (\pi/d_z)^2}$, with $m = \pm 1, \pm 3, \pm 5, \ldots$ for symmetric modes, and $m = \pm 2, \pm 4, \pm 6, \ldots$ for antisymmetric modes, where the sign of the *m* indicating the mode propagation direction. This results in a relatively simple expression for the Fano-like resonances in the angle-wavelength domain:

$$\lambda_m(\theta) \approx \frac{d_x n}{\sqrt{1 + \left(\frac{md_x}{2d_z}\right)^2}} \pm \frac{d_x \sin(\theta) sgn(m)}{1 + \left(\frac{md_x}{2d_z}\right)^2} \quad (1)$$

This provides the structure of resonances in the angle-wavelength domain. This consists of the left/right inclines resonances corresponding to the modes propagating to the right/left. These resonances cross at the angle-wavelength domain $\theta = 0$.

Fig.2. shows the resonance line (half of the resonance cross) for one lowest order mode m=1. The (1) reproduces the resonant line, however it says nothing about the depth, width, and in general the shape of the resonances. It is important to calculate the shape of the resonances in transmission and reflection. The modulated thin films however contain many parameters and have different architectures, which neither allow to derive simple analytical models for particular configurations, nor allow to derive universal models. A derivation of a simple universal relation, even an approximate one, but which would describe the main properties of such an interaction, would be satisfactory to develop an intuition of the waves propagating in such system and would allow us to predict the expected phenomena. Such a model was derived in [23], and gives the universal solution for reflection and transmission from such a film:

$$r = -\frac{c(\gamma - i(\omega - \omega_0))}{c + (2i + c)(\gamma - i(\omega - \omega_0))} \quad (2.a)$$

$$t = \frac{c + 2i\gamma + 2(\omega - \omega_0)}{c + (2i + c)(\gamma - i(\omega - \omega_0))} \quad (2.b)$$

The solutions contain generic parameters: c - the coupling coefficient between the incoming wave and waveguiding modes, $\gamma$ – the loss of the waveguiding mode (if $\gamma < 0$ then instead of losses there is a gain), and $\omega_0$ is the eigenfrequency of the waveguiding mode. The close-to-resonance assumption $|\omega - \omega_0| \ll \omega_0$ was considered. The resonance frequency in (2) relates with (1) by $\omega_0(\theta) = 2\pi c/\lambda_m(\theta)$. The (2) can be thought as the cross-section of the resonance by varying the wavelength, and fixing the incidence angle.

The analysis of (2) shows a shift of the resonance from that obtained by geometric consideration $\omega_0$. Interestingly, the reflection/transmission achieves large values (diverges) for sufficiently large negative values of $\gamma$. The divergence point is $\gamma_{div} = -c^2/(4 + c^2) \approx -c^2/4$ for $c \ll 1$. A simple estimation of maximal reflection at resonance can be obtained in the limit $c \ll 1$ and $|\gamma| \ll 1$:

$$r = -\frac{1}{1 + 4\gamma/c^2} \quad (3)$$

For $\gamma < \gamma_{div}$ the structure starts generating, i.e. the amplitudes of waveguiding modes start growing exponentially even without the injection. This follows from the autonomous part of the full solution. However, the other part of the solution, the driven one (2), remains bounded for $\gamma < \gamma_{div}$, except for a singularity at the $\gamma = \gamma_{div}$ point.

The width of the (shifted) resonance close to the singular point $\gamma_{div}$, $\gamma = \gamma_{div} + \Delta\gamma$: is given by a simple relation: $\Delta\omega \approx |\Delta\gamma|$, i.e. approaches zero at a singular (leasing) point.

**Numerical analysis:** The resonant amplifying behavior stems from the guided modes, that once leaked into an active substrate, are amplified. The guided mode corresponds to the Fano mode (in a particular example in Fig.1. it corresponds to the 1-st order Fano mode) which has a high resonant $Q$ parameter as well as a partial leakiness. The resonant amplification is realized when an input plane wave is amplified via an induced guided mode which is coupled with substrate with gain. The first-order diffraction of the incident wave propagates along the *x* direction once guided mode resonance (GMR) is attained. The key point of obtaining amplification in RCWA is via positively valued imaginary refractive index (Im $n > 0$), thus exponentially amplifying a guided mode along its direction of propagation. In this work we have chosen this value to be 0.007j, resulting in the amplification coefficient of 826.7 cm$^{-1}$. In the numerical model, the substrate complex refractive index remains constant for all wavelengths of incidence to evaluate amplified reflection characteristics of Fano resonance properly.

The coupling between the incident wave and waveguiding mode depends on the geometrical and material properties of the thin film: modulation depth $h$, thickness $L$, transverse periodicity $\Lambda_x$ and its refractive index vale $n_f$. Moreover, correct tuning of such parameters maximizes the reflection values in the parameter space. The parameters were chosen to be: $h = 100$ nm, $L = 87.5$ nm, $\Lambda_x = 700$ nm, $n_{film} = 2.24$ and $n_{sub} = 1.47 + 0.007$j for an entire spectrum, thus achieving amplification at normal incidence for 1064 nm. The results for such parameters are presented in Fig.2.

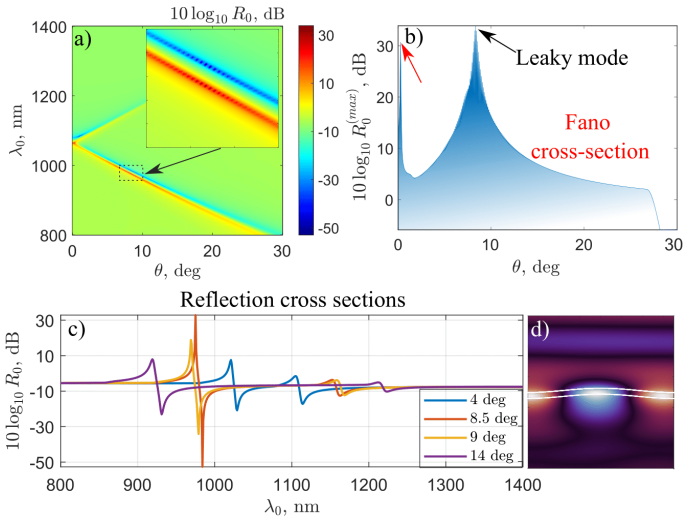

*Fig.2. a) Reflection map: the reflection coefficient in two-parameter spaces of the incidence angle and wavelength ($\theta, \lambda_0$). b) angular dependence of reflection peak value; c) reflection dependences on the wavelength for different incidence angles (vertical cross-sections of the reflection map). c) Reflection dependence on the incidence angle along the resonance (for resonant wavelength). The maximum increase of the resonance is observed at the angle $\theta = 9$ at $\lambda_0 = 970$ nm, whose field configuration is shown in d).*

The results correspond to an *s-polarized* input plane wave of incidence with a vacuum wavelength from $\lambda_0 = 400$ nm to from $\lambda_0 = 1400$ nm and angle of incidence form $\theta = 0$ deg to $\theta = 30$ deg. Judging by the calculated results, shown in Fig.2 (a,b), the leaky mode amplification at maximum obtains the value of 33.89 dB ($\theta = 8.35$ deg). Note, that at a normal incidence angle ($\theta = 0$ deg) the increase of reflection is observed as well, as the two resonances (related to right and left propagating modes) coalesce, however, the peak reflection value is 13.73 dB. At near normal incidence, however, it is 30.67 dB ($\theta = 0.25$ deg), hinting that leaky modes are best suited to amplify reflection.

To explore the effect, we performed the analysis in detail by parameter sweep. The sweep of the thickness of the film $L$, and modulation depth $d$ for the other parameter fixed, is presented in Fig. 3. The variation of the width of the film changes the wavenumber of propagation mode, i.e. the resonance condition. The resonances with the 1$^{st}$, 2$^{nd}$, and 3$^{rd}$ modes are visible.

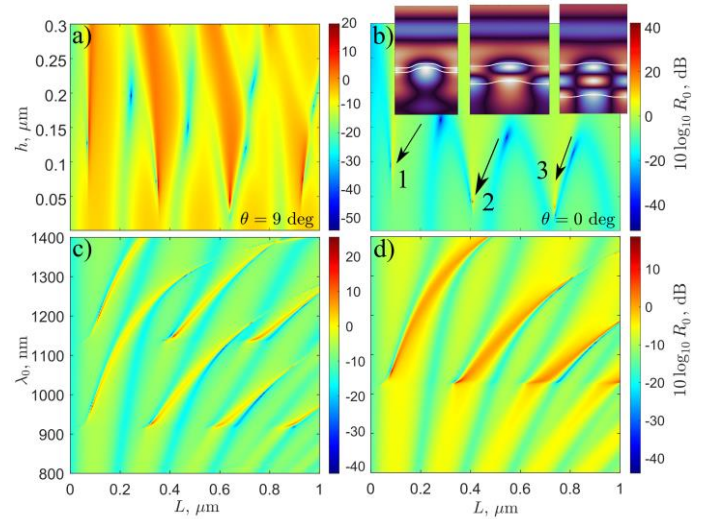

*Fig. 3. Geometrical parameter sweep of a meta-amplifier for isolated Fano resonances (946 nm) (a) and a coalescence of two Fano resonances (1064 nm) (b). Thickness parameter sweep of the GMR filter for $\theta = 9$ deg (c) and $\theta = 0$ deg (d).*

Those guided modes provide the strongest amplifications. The spectral analysis for an oblique angle case (see Fig. 3 (a)) reveals a split of Fano resonance, which creates additional leaky modes that are used to enhance reflection. The thickness sweep analysis (see Fig. 3 (c,d)) confirms that trend and illustrates the dependence of the reflection peak. A redshift of a peak reflection coefficient is observed due to the increase in thickness and the decrease of the peak reflection coefficient, thus implying that the resonance is detuned, and the structure likely stops lasing generation.

Moreover, peak reflection dependence on the imaginary part of the refractive index is illustrated in Fig. 4. (a) for a leaky mode and in Fig. 4. (b) for a conjunction of Fano resonances (normal incidence). The ideal coupling between leaky modes and active substrate is at the imaginary refractive index of $\gamma = 0.007$j. This tendency is observed for both oblique and normal incidence of an input wave and concludes that lasing is achievable for imaginary refractive index at the order of $10^{-3}$.

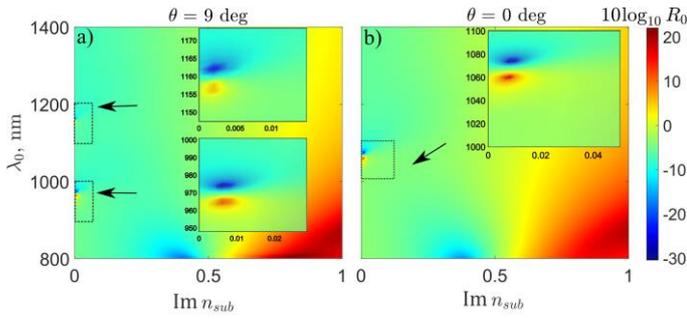

*Fig. 4. Amplification map in parameter space of the imaginary part of the substrate's refractive index $n_{sub}$, and input wavelength $\lambda_0$ a) case corresponds to the singular Fano mode and b) case corresponds to Fano crossing. All of the other geometrical parameters and material properties remain the same.*

**To conclude,** we provided numerical simulations of a Fano resonance in a modulated thin layered waveguide on the substrate that exhibits gain. Furthermore, we provide an analysis of resonance behavior at the Fano crossing effect as well as the leaky mode regime which provides a strong amplification of the reflected EM field. We also expand such an idea via a simple analytical model to explain the reasoning behind such phenomena. Analytical treatment of coupled oscillator model results in a lasing system once gain coefficient $\gamma$ reaches a large enough value, thus the reflection coefficient is divergent. Numerical simulations via RCWA confirm this via observed sharp peaks in reflection spectra and we identify them as lasing points. We verify that those sharp peaks operate in leaky mode regime and the highest reflection coefficient is at $\gamma = 0.007\,j$, i.e., power of $10^{-3}$, and first order Fano resonance. Such devices can be applied in micro-lasing systems, amplifiers and active spatial filters.

**Acknowledgements**

This work was supported by Spanish Ministry of Science, Innovation and Universities (MICINN) under the project PID2022-138321NB-C21, and by the ArtPM project from the Research Council of Lithuania (LMTLT), agreement No. S-MIP-24-10.